\documentclass{webofc}

\usepackage[varg]{txfonts}   
\usepackage{hyperref}
\usepackage{url}
\hypersetup{colorlinks=true,citecolor=blue,urlcolor=blue,linkcolor=blue}
\usepackage{xurl}
\usepackage{lipsum}
\usepackage{amsmath}
\usepackage{float}
\usepackage{subcaption}
\usepackage{graphicx}
\usepackage{tabularx}
\usepackage{soul}
\usepackage{geometry}
\usepackage[normalem]{ulem}
\usepackage[dvipsnames]{xcolor}
\usepackage{array}
\usepackage{mathtools}

\begin{document}
\title{A Double-Dilaton Holographic Model of QCD: Running Coupling and Meson Phenomenology}
%
%

\author{\firstname{Héctor} \lastname{Cancio}\inst{1}\fnsep\thanks{\email{hcancio@ifae.es}} \and
        \firstname{Pere} \lastname{Masjuan}\inst{1}\fnsep\thanks{\email{masjuan@ifae.es}}
}

\institute{Grup de F\'isica Te\`orica, Departament de F\'isica, Universitat Aut\`onoma de Barcelona, Institut de F\'isica d'Altes Energies (IFAE) and The Barcelona Institute of Science and Technology (BIST), 08193 Bellaterra (Barcelona), Spain}

\abstract{We present a holographic model of QCD to study the strong coupling constant $\alpha_s$ from the low-energy region to the high-energy regime. The approach uses a double-dilaton background, which breaks conformal and chiral symmetries, while large-$N_c$ corrections are incorporated to connect the holographic description with perturbative QCD. The model leads to an infrared fixed point and provides a smooth transition to the pQCD running. The resulting background also produces linear Regge trajectories for scalar, vector, and tensor states in reasonable agreement with experimental data.
}
\maketitle
\section{Introduction}
\label{Introduction}
The low-energy regime of QCD poses theoretical challenges and various approaches have been employed to address this region, with holography emerging as one of the most promising. Using the AdS/CFT correspondence \cite{Maldacena}, phenomenological models such as the Soft Wall model \cite{Karch} and its variations have been widely used to describe the meson spectrum, investigate Regge trajectories  and compute electormagnetic form factors \cite{CancioMasjuan, BallonBayona}, as well as hadronic form factors in general \cite{deTeramondBrodskyDosch}.

Our aim is to construct a model able to describe the running of the strong coupling constant $\alpha_s$ at low energies and find a matching with the high-energy result from perturbative QCD \cite{CancioMasjuan0}. In the past, QCD running was obtained from Holographic Light Front QCD (HLFQCD) \cite{Brodsky}. Other approaches include phenomenological information in both perturbative and non-perturbative regimes as in \cite{DeSanctis}. In our case, we will use the Ricci Flow equation \cite{Friedan}, \cite{Hamilton}, \cite{Perelman}, which describes the evolution of the metric tensor with the holographic coordiante, to describe the running of $\alpha_s$.

This letter is organized as follows. In Section \ref{Sec.RicciFlow}, we give the basics of Ricci Flow and its relation to the QCD running coupling constant $\alpha_s$. In Section \ref{Sec.DDSW}, we propose a new holographic QCD model, and we obtain its predictions: running of  $\alpha_s$ and radial Regge trajectories. Finally in Section \ref{Sec.Conclusions} we summarize our findings.
\section{Ricci Flow}
\label{Sec.RicciFlow}
We work in a AdS spacetime with the following metric tensor:
\begin{equation}
\label{eqn:AdSBackground}
ds^2=\frac{R^2}{z^2}\left(dz^2+\eta_{\mu\nu}dx^{\mu}dx^{\nu}\right).
\end{equation}
Here $R$ is the AdS radius, which is related to 't Hooft's coupling $\lambda$ by the formula $R^2/\alpha'=\sqrt{\lambda}$,
where $\lambda=g_{\text{YM}}^2N_c$. From here we have a relation between $g_{\text{YM}}$, the QCD strong coupling, and the geometrical data of AdS, namely the AdS radius $R$, so we can write:
\begin{equation}
\label{eqn:AdSCFTdictionary}
\alpha_s=\frac{R^4}{4\pi N_c \alpha'^2}.
\end{equation}
In the above equation $\alpha_s$ is constant, but we observe that a $z$-dependent AdS radius would give a running of $\alpha_s$, obtaining a deviation from AdS spacetime as background. This deviation can be interpreted as a dilaton field which breaks conformal symmetry \cite{Karch}, so assuming that the metric tensor depends on a renormalization scale $\mu$, its variation is given by (see \cite{CancioMasjuan0}):

\begin{equation}\label{eq:RGE2}
\mu \frac{\partial g_{\mu\nu}}{\partial \mu}
=2g_{\mu\nu}(\mu)\Bigl(1+\frac{\mu}{4}\frac{d}{d\mu}\log (\alpha_s(\mu^2))\Bigr).
\end{equation}
In this way, we can identify the second term $\beta_{\mu\nu}=g_{\mu\nu}\frac{\mu}{4}\frac{d}{d\mu}\log \alpha_s$ as an homologous to a $\beta$-function of the metric $g_{\mu\nu}$, and relate naturally the running of $\alpha_s(\mu^2)$ with the running of the metric tensor $g_{\mu\nu}(\mu)$. Introducing $t=\log{\mu}$, it follows the so-called Ricci Flow equation \cite{Friedan, Hamilton, Perelman}:
\begin{equation}
\label{RicciFlow}
\frac{\partial g_{\mu\nu}}{\partial t}=2g_{\mu\nu}-2R_{\mu\nu}.
\end{equation}


\section{Double Dilaton Soft Wall Model and its predictions}
\label{Sec.DDSW}
\subsection{Running of $\alpha_s$}
\begin{figure}[h]
\centering
\includegraphics[scale=0.45]{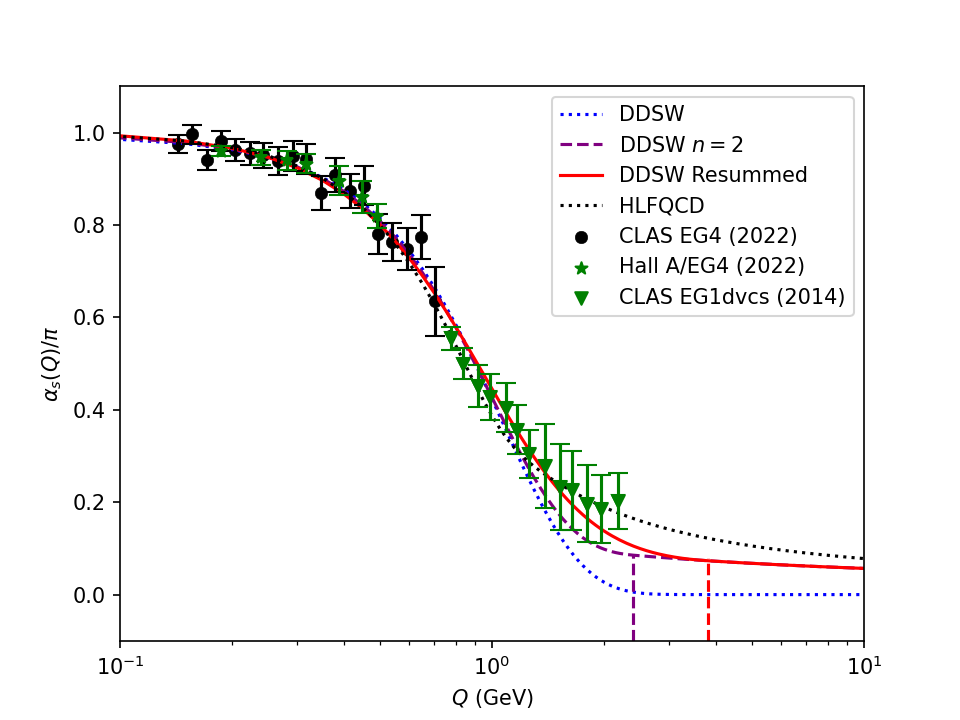}
\caption{Comparison of the model from Eq.~(\ref{eqn:parametrization}) (dotted blue) with the perturbative QCD running coupling obtained by matching to the DDSW model Eq.~(\ref{eqn:largeN}), for $n=2$ (dashed purple), and with its resummed version, Eq.~(\ref{eqn:Resum}) (red). The corresponding $\chi^2/\mathrm{DOF}$ values are $1.45$, $0.99$, and $0.93$, respectively. For comparison, we also show the HLF model from Ref.~\cite{Brodsky} (dotted black). The vertical dashed lines mark the matching points to perturbative QCD for the $n=2$ and resummed cases.
}
\label{figure:running}
\end{figure}

By evolving the metric tensor with Eq.(\ref{RicciFlow}) we can explore possible dilaton fields and its associated running coupling constant. A confining dilaton was used in \cite{Brodsky}, which had the opposite sign in comparison with the original dilaton presented in \cite{Karch}. In Ref.\cite{CancioMasjuan} we introduced a model with two dilatons, one with a positive and one with a negative sign, which enables for breaking the chiral symmetry and to study the running of $\alpha_s$ at the same time. Physically, this setup should correspond to a system with vector and axial-vector mesons, which can see different geometries. We call this set-up Double Dilaton Soft Wall model (DDSW).
Dilatons with different signs were used in the past in order to study its confinement properties and axial-vector mesons (see, for example, \cite{Afonin}), our approach, however, has the advantage of avoiding the need to choose a single dilaton sign to describe vectors and axial-vectors simultaneously.
From this set-up the analytic expression for the running of $\alpha_s$ predicted by the DDSW model is \cite{CancioMasjuan}:
\begin{equation}
\label{eqn:parametrization}
\frac{\alpha_s(Q)}{\pi}=\frac{a}{\cosh^{4/5}{(b(Q^2+1))}},
\end{equation}
where $a$ and $b$ are constants to be fitted with experimental data \cite{Deur} obtained from JLab experiments Hall CLAS EG4 (from Q = 0.143 GeV to 0.704 GeV), CLAS EG4/E977110 (from Q = 0.187 GeV to 0.490 GeV) and EG1dvcs (from $Q=0.775 \text{ GeV}$ to $2.177 \text{ GeV}$), depicted in Fig.\ref{figure:running} as solid star, solid circle, and solid triangle, respectively. We have normalized the function at the fixed-point $\alpha_s(Q \simeq 0)/\pi \simeq 1$ in Fig.\ref{figure:running}. From a fit to data with the model in Eq.(\ref{eqn:parametrization}), we obtain $a = 1.545 \pm 0.047$, $b=1.150 \pm  0.041 \text{ GeV}^{-2}$ with $\chi^2/DOF = 1.45$. Although the fit provides a reasonably good description of the data, it does not reproduce the pQCD running of the coupling constant at high energies. This behavior is not unexpected, as holographic QCD models are only applicable in the strongly coupled regime. At sufficiently high energies, the coupling is expected to smoothly transition towards the perturbative behavior predicted by pQCD.

Our running has been deduced from a holographic model based on AdS/CFT duality, so in order to reproduce physical data, one should expect large-$N_c$ corrections to be important. In Ref.\cite{CancioMasjuan0} we included them, for the first time, yielding the following result:
\begin{equation}
\label{eqn:largeN}
\frac{\alpha_s(Q)}{\pi}=c\sum_{g=0}^{n-1} \left(\frac{a}{\cosh^{4/5}{(b(Q^2+1))}}\right)^g.
\end{equation}
The parameters $a,b$ must be determined from data while $c$ is determined by imposing a matching point $Q_0$ to pQCD, in particular we must solve a system of two equations with unknowns $c$ and $Q_0$: 
\begin{equation}
\label{eqn:matchingEq}
\alpha_s^{\text{DDSW}}(Q_0)=\alpha_s^{\text{pQCD}}(Q_0),\quad
\frac{d\alpha_s^{\text{DDSW}}}{dQ}\bigg{|}_{Q=Q_0}=\frac{d\alpha_s^{\text{pQCD}}}{dQ}\bigg{|}_{Q=Q_0}.
\end{equation}
Solving the system recursively, starting with the initial value $c=\frac{\alpha_s(M_z)}{\pi} = \frac{0.118}{\pi} \simeq 0.0376$ and fitting the data, we obtain $a$ and $b$. By solving the system we determine $Q_0$.
For $n=2$, we obtain, after fitting to data, $a=18.87 \pm 0.60$, $b=(1.289 \pm 0.041) \text{ GeV}^{-2}$, $c=0.082$ and a $\chi^2/DOF = 0.99$. This result is shown in Fig.\ref{figure:running} as a purple-dashed line (notice we plot $\alpha_s(Q)/\pi$), with a matching with pQCD at 2.39 GeV.

Returning to Eq.(\ref{eqn:largeN}), it has de form of a geometric series that can be resummed (suggested by the fitted coefficients) giving a new matching point $Q=3.79 \text{ GeV}$ (see Fig.\ref{figure:running}), obtaining:
\begin{equation}
\label{eqn:Resum}
\frac{\alpha_s(Q)}{\pi}=\frac{c}{1-a \text{ sech}^{4/5}(b(Q^2+1))}.
\end{equation}

\subsection{Radial Regge trajectories}
\begin{figure}[H]
\centering
\begin{subfigure}[b]{0.32\textwidth}
\centering
\includegraphics[width=\textwidth]{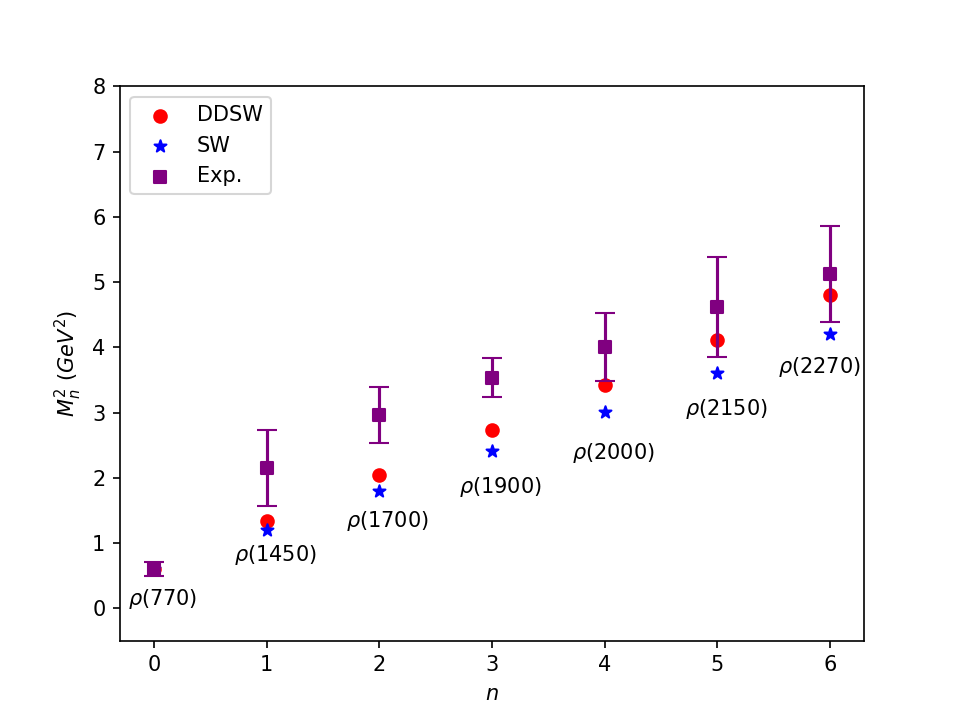}
\caption{Vector mesons}
\label{figure:regge}
\end{subfigure}
\hfill
\begin{subfigure}[b]{0.32\textwidth}
\centering
\includegraphics[width=\textwidth]{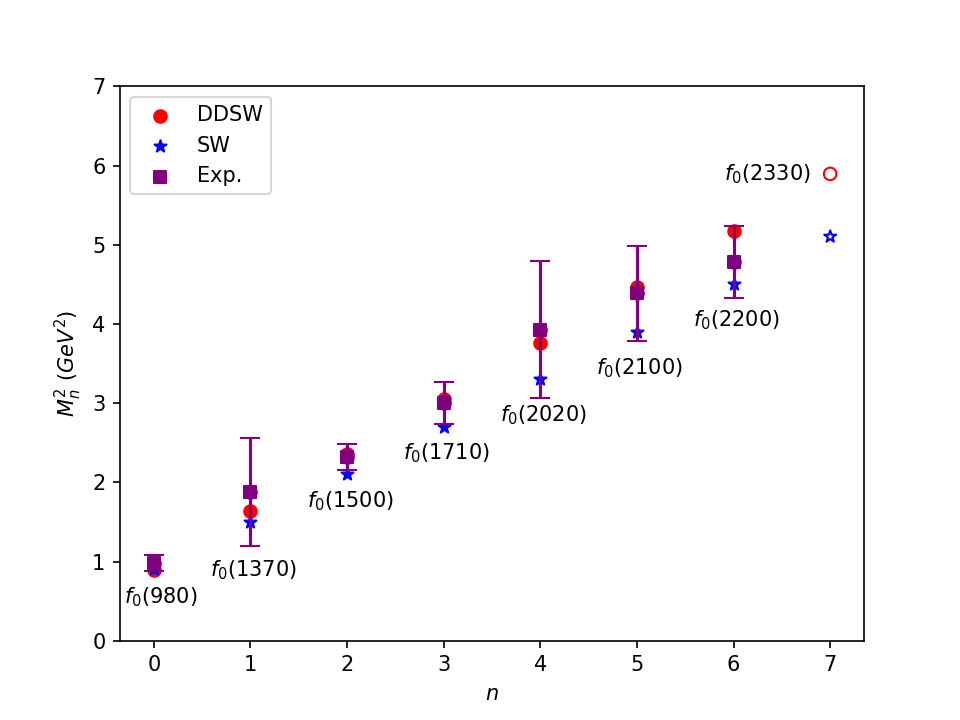}
\caption{Scalar mesons}
\label{figure:reggeScalar}
\end{subfigure}
\hfill
\begin{subfigure}[b]{0.32\textwidth}
\centering
\includegraphics[width=\textwidth]{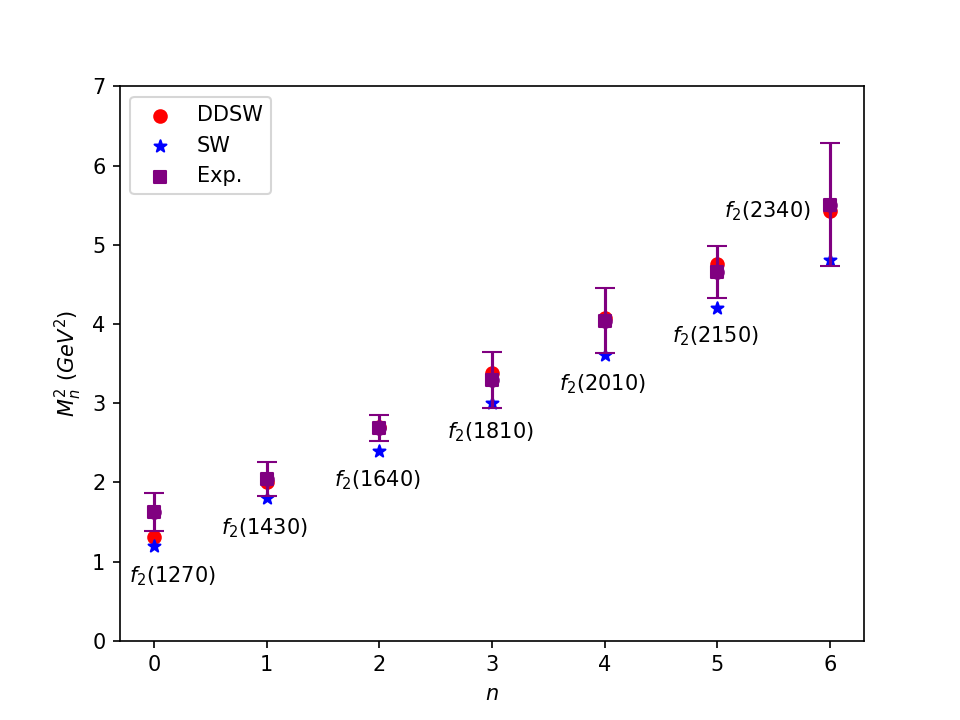}
\caption{Tensor mesons}
\label{figure:reggeTensor}
\end{subfigure}
\caption{Radial Regge trajectories $(M_n^2,n)$ for vector (left), scalar (center), and tensor (right) mesons in the DDSW model (red points), compared with the corresponding SW model predictions (blue stars) and available experimental data \cite{Navas} (purple boxes, we have used the half-width rule \cite{Masjuan:2012gc}). For the tensor case, the trajectory represents the proposed DDSW ordering along the Regge trajectory.}
\label{figure:reggeAll}
\end{figure}

Since we have a holographic model, we can study the corresponding equation of motion in order to obtain radial Regge trajectories. To do so, we fix the $\rho$-meson mass to $M_\rho=775 \text{ MeV}$ following \cite{Karch}. As a result we obtain $M_n^2\sim n$, that is, radial Regge trajectories for DDSW model are linear (as in the Soft Wall model).
We obtained also the spectrum of scalar and tensor mesons. An improvement respect the original Soft Wall model is achieved, and more accurate results are obtained when comparing to experimental data \cite{Navas}. In fact,  scalar resonances with $n=2, 3, 5$ show a departure to the experimental counterpart below $1\%$. Errors
are obtained via the half-width rule (i.e, $M_n^2\pm M_n\Gamma_n$, 
with $\Gamma_n$ the total decay width of the particle \cite{Masjuan:2012gc}). In Figure \ref{figure:reggeScalar}, the empty circle and the empty star represent the prediction of DDSW and Soft Wall models, respectively, for $f_0(2330)$ meson.

\section{Conclusions}
\label{Sec.Conclusions}
In this work, we have introduced the Double Dilaton Soft Wall (DDSW) model as a holographic framework for describing the QCD coupling at all energies. The model provides an infrared fixed point and can be matched to the perturbative QCD running at high energies through including for the first time large-$N_c$ corrections.

We have also investigated the meson spectrum within the DDSW background. Linear Regge trajectories are obtained for scalar, vector, and tensor mesons, with the scalar and vector spectra showing improved agreement with experiment compared with the original Soft Wall model.
These results suggest that the DDSW framework can be extended to other holographic backgrounds and be further exploit in calculations where a robust description of $\alpha_s$ is necessary.

\section{Acknowledgements}
Our work has been supported by the Ministry of Science and Innovation (MICINN) under grant PID2023-146142NB-I00, and through the State Research Agency under the Severo Ochoa Centres of Excellence Programme 2025–2029 (CEX2024-001441-S). IFAE is partially funded by the CERCA program of the Generalitat de Catalunya. 


%
%
%

\end{document}